\documentclass[a4paper,11pt]{article}

\usepackage[dvipsnames]{xcolor}
\usepackage{jheppub}
\usepackage{graphicx}
\usepackage{amsmath,amssymb,amsthm,amsfonts}
\usepackage{slashed}
\usepackage{braket}
\usepackage[export]{adjustbox}
\usepackage{enumitem}
\usepackage{placeins}
\usepackage[normalem]{ulem}
\usepackage{multirow}
\usepackage{hyperref}
\usepackage{mathrsfs}  
\usepackage{dsfont}

\usepackage{verbatim}
\usepackage[T1]{fontenc}
\usepackage[utf8]{inputenc}
\usepackage{array}
\usepackage{wrapfig}
\usepackage{tabu}
\usepackage{longtable}
\usepackage{float}
\usepackage{caption}
\usepackage{subcaption}
\usepackage{tikz,tikz-3dplot}
\usepackage[compat=1.0.0]{tikz-feynman}
\usetikzlibrary{calc,arrows.meta,shapes.geometric}
\usepackage{soul}
\usepackage[usestackEOL]{stackengine}
\usepackage{cancel}

\allowdisplaybreaks

\definecolor{mycolor}{rgb}{0.015,0.386, 0.0273}
\hypersetup{
    colorlinks=true,
    linkcolor=mycolor,
    citecolor=mycolor,
    filecolor=mycolor,      
    urlcolor=mycolor
    }

\def\beq{\begin{equation}}
\def\eeq{\end{equation}}
\def\bea{\begin{eqnarray}}
\def\eea{\end{eqnarray}}

\preprint{SLAC-PUB-17781}

\title{Splitting amplitudes at N$^3$LO in QCD}

\author[a]{Xin Guan,}
\emailAdd{guanxin@slac.stanford.edu}
\author[b]{Franz Herzog,}
\emailAdd{fherzog@ed.ac.uk}
\author[c]{Yao Ma,}
\emailAdd{yaomay@phys.ethz.ch}
\author[a]{Bernhard Mistlberger,}
\emailAdd{bernhard.mistlberger@gmail.com}
\author[a]{Adi Suresh}
\emailAdd{adisur@stanford.edu}

\affiliation[a]{SLAC National Accelerator Laboratory, Stanford University, Stanford, CA 94039, USA}
\affiliation[b]{Higgs Centre for Theoretical Physics, School of Physics and Astronomy,\\The University of Edinburgh, Edinburgh EH9 3FD, Scotland, UK}
\affiliation[c]{Institute for Theoretical Physics, ETH Zürich, 8093 Zürich, Switzerland}

\abstract{
In the limit where partons become collinear to each other, scattering amplitudes factorize into a product of universal, process-independent building blocks and scattering amplitudes involving fewer partons. We compute these universal building blocks---known as splitting amplitudes---for two collinear QCD partons up to third loop order in QCD. Our results describe arbitrary time-like splitting processes. Due to the violation of strict collinear factorization in space-like splitting processes, we specifically present space-like splitting amplitudes for three-parton QCD scattering amplitudes at third loop order. To achieve our results, we perform a collinear expansion of three-loop scattering amplitudes using a new expansion-by-subgraph technology, which is based on the method of regions.
}

\begin{document}
\maketitle

\section{Introduction}

% Lance and friedns computed the $g\to gg$ splitting amplitudes at 2 loops in ref.~\cite{Bern:2004cz}.
% Simon and Nigel for $q\to q g$ in ref.~\cite{Badger:2004uk}.

An important property of gauge theory scattering amplitudes is that they factorize in soft and collinear (IRC) limits of the external states. 
Factorization implies that a scattering amplitude in these limits is given by a product of lower multiplicity amplitudes times a universal, process-independent singular function. 
These kinematic limits are of great interest for several reasons. 
For one, partonic cross sections for fixed parton multiplicity are singular in these limits.
In the case of colorless initial states, e.g. $e^+e^-$, IR divergences cancel in the sum over final states  contributing to the same IRC-safe observable~\cite{Kinoshita:1962ur,Lee:1964is}.
In the case of hadronic initial states, such as the proton-proton collisions at the LHC, initial-state collinear singularities need to be absorbed into a redefinition of parton distributions functions (PDFs)~\cite{Politzer:1974fr,Georgi:1974wnj,Altarelli:1977zs}.
In practice, handling the IRC divergences in fully exclusive processes requires the introduction of infrared subtraction (including slicing) schemes to render the individual partonic cross sections finite. By providing the required building blocks, IRC limits of amplitudes have been the backbone of such subtraction schemes at both next-to-leading order (NLO)~\cite{Catani:1996vz,Frixione:1995ms} and next-to-NLO (NNLO) in QCD~\cite{Ridder2005, Somogyi:2006da, Somogyi:2006db, Catani:2007vq, Czakon:2010td, Boughezal:2011jf, Gehrmann_De_Ridder_2012, Gaunt:2015pea, Caola:2017dug, Anastasiou:2018rib, Herzog:2018ily, Magnea:2018hab, Cieri:2018oms, Anastasiou:2022eym}. 
Furthermore, IRC limits are of interest as they provide universal building blocks for resummation formulae~\cite{Collins:1989gx,Sterman:1995fz,Bauer:2000ew,Bauer:2001yt}. They also provide valuable constraints and useful data for the analytic reconstruction program, see for example refs.~\cite{Bern:1993qk,Almelid:2017qju,Caron-Huot:2019vjl,Dixon:2023kop,Caron-Huot:2020bkp,Dixon:2022rse}, and they serve as an excellent cross-check when new scattering amplitudes are computed.
 
In QCD the current status of the art is that single soft limits are known up to two loops for the general case \cite{Weinberg:1965nx,Bern:1995ix,Bern:1998sc,Bern:1999ry,Catani:2000pi,Badger:2004uk,Li:2013lsa,Duhr:2013msa,Dixon:2019lnw}, and up to three loops in the case of only 2 colored external partons. Double soft limits are known to one loop~\cite{Catani:1999ss,Zhu:2020ftr,Czakon:2022dwk,Catani:2021kcy}, and triple soft limits are known at tree level~\cite{Catani:2019nqv,DelDuca:2022noh,Catani:2022hkb}.

The main subject of this work is the collinear limit, by which we mean the limit in which two external partons become collinear.
In this scenario an $n$-point amplitude factorizes into a product of a universal splitting amplitudes, containing the singular behaviour, times an $(n-1)$-point amplitude. 
For squared or interfered amplitudes singular behaviour in the collinear limit is captured by the splitting functions. 
In general, due to spin correlations, splitting functions have tensorial structure in the Lorentz or spinor indices; see, e.g., ref.~\cite{Catani:1998nv}.
At the level of helicity amplitudes, this spin correlation can be resolved as a factorization into helicity-dependent splitting amplitudes, which can be seen as simpler building blocks of splitting functions. 
General properties of collinear factorization at the amplitude level were studied in refs.~\cite{Kosower:1999xi, Bern:1994zx}. 
One-loop results for the splitting amplitude were extracted in refs.~\cite{Bern:1998sc, Bern:1999ry, Kosower:1999rx}. 
Two-loop results were presented for the $g\to gg$ case in ref.~\cite{Bern:2004cz} and for all channels in ref.~\cite{Badger:2004uk}. 
The two-loop splitting function at higher orders in the dimensional regulator was  presented in ref.~\cite{Duhr:2014nda}. 
We note here also that, in contrast to the time-like splitting, strict collinear factorization is violated in the space-like splitting at the amplitude level~\cite{Cieri:2024ytf,Catani:2011st,Catani:2012iw,Forshaw:2012bi,Kyrieleis:2006fh,Forshaw:2006fk,Schwartz:2017nmr}, i.e. when one of the collinear partons is in the initial state and the other is in the final state. 
While such factorization violating effects appear to cancel at the level of  cross sections in pure QCD up to NNLO, their impact at higher orders is still an open question~\cite{Forshaw:2012bi,Henn:2024qjq}.

For three collinear partons, the splitting function/amplitude is known up to one loop~\cite{Catani:1998nv, Catani:2003vu, Badger:2015cxa}, and for four it is known at tree level~\cite{ DelDuca:2019ggv,DelDuca:2020vst}. 
In this article, we will take one step further by computing the three-loop splitting amplitude for two collinear partons.

While in the context of IRC subtraction only required at next-to-next-to-NNLO (N$^4$LO) the three-loop splitting amplitudes also provide the key ingredient for the real triple-virtual correction to the next-to-NNLO (N$^3$LO) Altarelli-Parisi (AP) splitting kernels (the anomalous dimensions of the PDFs) in the approach developed in ref.~\cite{Kosower:2003np}. 
While considerable progress has been made towards an exact determination, the AP splitting kernels are so far known only approximately at N$^3$LO \cite{Moch:2017uml, Falcioni:2023vqq, Falcioni:2023luc, Moch:2023tdj, Moch:2021qrk, Falcioni:2023tzp, Gehrmann:2023cqm, Gehrmann:2023iah}. 
An exact determination would still be highly desirable. 
The N$^3$LO splitting functions are essential not only for the determination of factorization scale uncertainties for N$^3$LO cross sections, such as Higgs production~\cite{Anastasiou:2015ema,Mistlberger:2018etf,Duhr:2019kwi} and the Drell-Yan process \cite{Duhr:2020seh,Duhr:2020sdp}, but also for the determination of the N$^3$LO PDFs themselves~\cite{NNPDF:2024nan,Cridge:2024exf,Cridge:2023ryv,Cooper-Sarkar:2024crx}.

To facilitate the computation of the three-loop splitting amplitudes, we will use a novel method to extract the collinear limit from a set of full kinematic amplitudes. 
The difference to former amplitude extractions of splitting amplitudes, such as the one by Glover and Badger \cite{Badger:2004uk} is that we instead take the collinear limit before loop integration at the level of the Feynman integrand. 
The well known method to accomplish this is the method of regions \cite{Smirnov:1990rz,Smirnov:1994tg,Beneke:1997zp}, which we implement in momentum space via a novel correspondence between subgraphs of the Feynman diagrams and the set of contributing regions. 
This approach was first developed in the context of the on-shell expansion \cite{Gardi:2022khw}, before it was employed by some of the authors in a recent calculation of the three-loop soft limit \cite{Herzog:2023sgb}. 
A more rigorous work was also provided by one of us, justifying the procedure in a wider class of applications, with the possible inclusion of soft external momenta~\cite{Ma23}.

In this manner, we will obtain a momentum-space representation for the collinear limit of several Higgs decay amplitudes, from which we can extract the relevant splitting amplitudes.
To reduce the amplitude to a set of master integrals, we employ two independent implementations of Laporta's algorithm, implementing integration-by-parts (IBP) reduction~\cite{Laporta:2001dd,Chetyrkin1981,Tkachov1981}.
Notably, the {\tt Blade} package, which implements the block-triangular form method~\cite{Liu:2018dmc,Guan:2019bcx} for improved reduction, is publicly available~\cite{Guan:2024byi}.
The master integrals are subsequently brought into a canonical form~\cite{Henn:2013pwa} using algorithmic methods~\cite{Lee:2016bib} and solved using the method of differential equations~\cite{Gehrmann:1999as,Kotikov:1990kg,Kotikov:1991hm,Kotikov:1991pm} in terms of harmonic polylogarithms~\cite{Remiddi:1999ew,Maitre:2005uu}.
All but one very simple boundary condition for the solution of the differential equations are determined from consistency and regularity conditions~\cite{Henn:2020lye,Dulat:2014mda,Henn:2013nsa}.

Besides providing the first calculation of the three-loop splitting function, this work also presents the first application of the subgraph expansion formalism for the collinear expansion, which can, in principle, be developed to arbitrary loop orders with this methodology.
The methods for collinear expansion of Feynman integrands we employ here have previously been studied for Higgs and color-singlet production~\cite{Ebert:2020lxs}. They were used to derive N$^3$LO beam functions and fragmentation functions as well as to produce approximations of hadronic cross sections~\cite{Ebert:2020qef,Ebert:2020lxs,Ebert:2020unb,Ebert:2020yqt}.
The technique of identifying suitable regions and expanding Feynman integrands for collinear expansions developed in this article will allow us to extend these results to scattering amplitudes and cross sections involving arbitrarily many loop integrations.

This paper is structured as follows. The general setup is introduced in section~\ref{sec:setup}. The collinear limit and its associated expansion-by-subgraph is presented in section~\ref{sec:collinear}.
Details of the calculation and results for splitting amplitudes in QCD and $\mathcal{N}=4$ sYM theory are presented in section~\ref{sec:splittings}. Finally, we draw our conclusions in section~\ref{sec:conclusions}.
%\newpage
\section{Setup} 
%\subsection{Setup}
\label{sec:setup}
A massless QCD parton can be characterized by its flavour $f$, momentum $p$, color $c$, and helicity~$\lambda$. 
We denote the $n$-particle scattering amplitude by 
\beq
\mathcal{A}_{n}^{c_1\dots c_n}(\{p_1,f_1,\lambda_1\},\dots,\{p_n,f_n,\lambda_n\}).
\eeq
Fermions and anti-fermions are charged under the (anti-)fundamental representation of $SU(n_c)$, and gluons are under the adjoint color representation.
The color indices on our $n$-parton scattering amplitudes  consequently belong to different representations.
We maintain this notation here for illustrative purposes and will clarify further below.
A generic $n$-parton scattering amplitude can be expanded in perturbation theory as follows:
\begin{eqnarray}
    &&\mathcal{A}_{n}^{c_1\dots c_n}(\{p_1,f_1,\lambda_1\},\dots,\{p_n,f_n,\lambda_n\})\nonumber\\
    &&\qquad\qquad =g_S^{n-2}\sum_{o=0}^\infty a_S^{o}\ \mathcal{A}_{n}^{(o),\,c_1\dots c_n}(\{p_1,f_1,\lambda_1\},\dots,\{p_n,f_n,\lambda_n\}),
\end{eqnarray}
where we have introduced the parameter $a_S$, which is related to the strong coupling constant $\alpha_S$ via
\beq
\label{eq:asdef}
a_S=\frac{\alpha_S}{\pi}.
\eeq

\begin{figure}[!h]
  \begin{center}
  \includegraphics[scale=0.19]{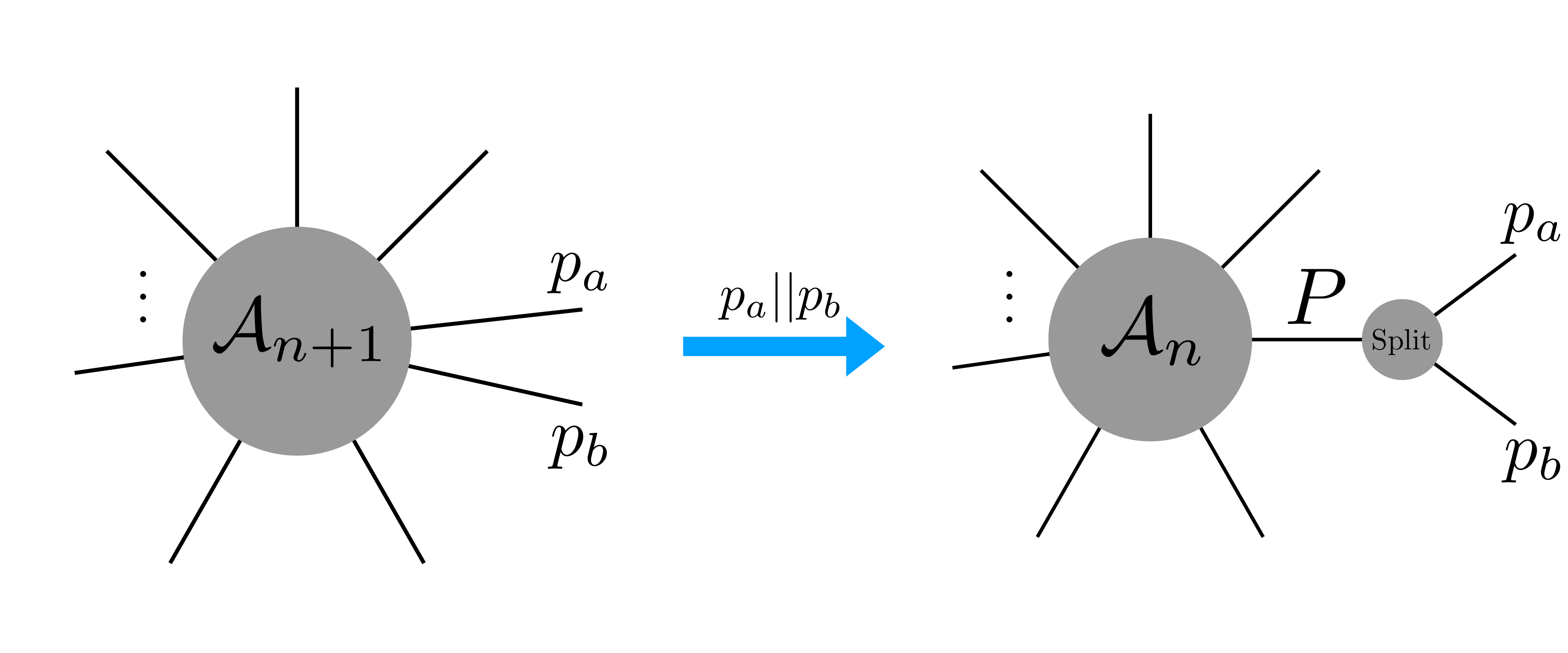}
  \caption{\label{fig:split} Schematic depiction of the 1$\to$2 splitting process.}
  \end{center}
\end{figure}
In this article, we consider the kinematic limit of QCD amplitudes in which two partons with momenta $p_a$ and $p_b$ become collinear to one another. 
Schematically, this process is depicted in figure~\ref{fig:split}.
We refer to the sum of the splitting momenta as $P$  such that
\beq
P=p_a+p_b.
\eeq
In the collinear limit (indicated by arrows below) we have
\beq
p_a\to z \tilde P,\hspace{1cm}p_b\to (1-z) \tilde P,\hspace{1cm} p_a+p_b=P\to\tilde P,
\eeq
where $z$ is the fraction of the light-like momentum $\tilde P$ carried by $p_a$ in the collinear limit.
Factorization of scattering amplitudes in the collinear limit then implies the following formula:
\begin{eqnarray}
&&\lim\limits_{a||b} \mathcal{A}_{n+1}^{c_1,\dots,c_a,c_b,c_{n+1}}(1,\dots,a,b,\dots,n+1)\nonumber\\
&&\qquad \qquad =\sum\limits_{\lambda_P=\pm} \text{Split}^{c_Pc_a c_b}_{-\lambda_P}(z;a,b)\ \mathcal{A}_{n}^{c_1,\dots,c_P,\dots,c_{n+1}}(1,\dots,\tilde P ,\dots,n+1).
\label{eq:splitformula}
\end{eqnarray}
%\YM{Should the $\lambda$ above be $\lambda_P$? Same for (2.7).}
Above, we used a short hand notation in the argument of the amplitude $i \sim \{p_i,f_i,\lambda_i\}$.
On the right-hand side of the above equation, there is a sum of the product of the $n$-parton amplitude and the splitting amplitude $\text{Split}^{c_P}_{-\lambda_P}$, over the positive and negative helicity components of the intermediate particle with momentum $P$.
We assume that the explicit splitting theorem above is valid in a pseudo-Euclidean scattering region and for time-like splitting processes~\cite{Catani:2011st}.
Analytic continuation and factorization in physical scattering regions will be discussed below. Violations of strict collinear factorization was also discussed in refs.~\cite{Cieri:2024ytf,Catani:2011st,Catani:2012iw,Forshaw:2012bi,Kyrieleis:2006fh,Forshaw:2006fk,Schwartz:2017nmr}.

The splitting amplitude itself can be expanded perturbatively.
\beq
\text{Split}^{c_Pc_a c_b}_{-\lambda_P}(z;a,b)=g_S \sum_{o=0}^\infty a_S^o\ \text{Split}^{(o),\, c_Pc_ac_b}_{-\lambda_P}(z;a,b).
\eeq
There are three distinct parton configurations possible in QCD.
\bea
g&\to& g g\ ,\nonumber\\
g&\to& q\bar q\ ,\\
q&\to& qg\ .\nonumber
\eea
Splitting amplitudes for anti-quarks or swapped final particles are related by symmetry.
However, all possible helicity configurations have to be considered.

In this article, we derive perturbative QCD corrections to all splitting amplitudes through third order, or N$^3$LO.
We derive these splitting amplitudes by taking the collinear limit of QCD scattering amplitudes of a Higgs boson and three partons. 
In particular, we consider the scattering amplitudes involving a Higgs boson and three gluons as well as a Higgs boson, a gluon, and a quark-anti-quark pair. 
We compute these two scattering amplitudes in five massless flavor QCD  and integrate out  the degrees of freedom of the top quark.
The Higgs boson then couples directly to gluons via an effective interaction~\cite{Wilczek1977,Shifman1978,Inami1983,Spiridonov:1988md}.
Furthermore, we compute a scattering amplitude of a Higgs boson, a gluon, and a bottom-anti-bottom quark pair. 
We construct this amplitude such that the Higgs boson couples to the bottom quark via a Yukawa interaction but treat the bottom quark otherwise as massless, see, for example, refs.~\cite{Dicus:1998hs,Balazs:1998sb,Harlander:2003ai,Duhr:2019kwi} for the application of the massless bottom quark limit to LHC cross sections.

Let us denote the three aforementioned amplitudes by
\beq
\label{eq:amplist}
\{\mathcal{A}_{hggg},\hspace{1cm}
\mathcal{A}_{hqg\bar q},\hspace{1cm}
\mathcal{A}_{hb\bar bg}\}.
\eeq
These amplitudes have been computed through two loops in refs.~\cite{Gehrmann:2023etk,Gehrmann:2011aa,Ahmed:2014pka} and a complete three-loop computation is yet missing.
A first result for a similar amplitude for an off-shell vector boson and three partons in the planar limit has become available in ref.~\cite{Gehrmann:2023jyv}.
To extract the splitting amplitudes, we take the strict collinear limit~\cite{Ebert:2020lxs}, which reduces the amplitude on the right-hand side of eq.~\eqref{eq:splitformula} to the tree-level amplitude.
By selecting specific helicity configurations, we can extract the splitting amplitudes.
Expanding the loop integrals around the strict collinear limit can be achieved by using the method of regions.
We shall demonstrate these constructions in the next section.
\section{The collinear limit}
\label{sec:collinear}

%\subsection{Parameterizing the collinear limit}
\subsection{Parameterization}
\label{sec:Parameterizing}
To distinguish three-parton amplitude momenta and two-parton amplitude momenta we denote the latter ones by $\tilde p_i$.
To develop our parameterization, we choose a specific frame:
\beq
\tilde{p}_1^\mu=\frac{\sqrt{s}}{2} \left( \begin{array}{c}1 \\ 0 \\ \vdots \\ -1 \end{array}\right), \hspace{1cm}
\tilde{p}_2^\mu=\frac{\sqrt{s}}{2} \left( \begin{array}{c}1 \\ 0 \\ \vdots \\ 1 \end{array}\right), \hspace{1cm }
s=(\tilde p_1+\tilde p_2)^2.
\eeq
For the three-parton amplitude we study here, we leave $p_2=\tilde p_2$ and consider the limit where $p_1 \parallel p_3$.
We further parameterize $p_1$ and $p_3$ via
\beq
p_1= z \tilde p_1 -\frac{k_\perp^2}{zs} \lambda \tilde p_2 +\sqrt{\lambda} k_\perp,\hspace{1cm }
p_3= (1-z) \tilde p_1 -\frac{k_\perp^2}{(1-z)s} \lambda \tilde p_2 -\sqrt{\lambda} k_\perp,\hspace{1cm }
k_\perp= \left( \begin{array}{c}0 \\ |k_\perp| \\0  \\ \vdots \\  0\end{array}\right).
\eeq
We introduced a small parameter $\lambda$ and, in the limit $\lambda \to 0$, the two momenta $p_1$ and $p_3$ become collinear.
(Note that this $\lambda$ is not the helicity.) With this parameterization, the three parton invariants are
\beq
s_{12}=z s,\hspace{1cm}
s_{23}=(1-z)s, \hspace{1cm}
s_{13}=-\lambda \frac{k_\perp^2}{z(1-z)}.
\eeq
The corresponding spinor brackets (see for example refs.~\cite{Bern:1991aq,Bern:1993mq,Dixon:2013uaa}) behave as follows in the collinear limit:
\bea
\langle 1| 2\rangle &=&\langle \tilde 1| \tilde 2\rangle  \sqrt{z}\sim\mathcal{O}(1),\hspace{1cm}
\langle 2| 3\rangle = \langle \tilde 1 | \tilde 2\rangle  \sqrt{1-z}\sim\mathcal{O}(1),\hspace{1cm}
\langle 1| 3\rangle \sim \mathcal{O}(\sqrt{\lambda}),\nonumber\\
\left[ 1 | 2 \right] &=&[ \tilde 1 |  \tilde 2 ]  \sqrt{z}\sim\mathcal{O}(1),\hspace{1cm}
[ 2 | 3 ] = [ \tilde 1| \tilde 2 ]  \sqrt{1-z}\sim\mathcal{O}(1),\hspace{1cm}
[ 1 | 3 ] \sim \mathcal{O}(\sqrt{\lambda}).
\eea
Above, the spinor brackets represent positive- and negative-helicity Weyl spinors $|i^\pm\rangle$, with $\langle i | j \rangle =\langle i^- | j^+ \rangle $ and $[ i | j ] =\langle i^+ | j^- \rangle $.

In the case of a ``parent'' parton with momentum $\tilde p_1$ splitting into two collinear partons with momenta $p_1$ and $p_3$, the variable $z$ describes the fraction of $\tilde p_1$ carried by the particle with momentum $p_1$.
To describe the case where an initial state parton with momentum $p_1$ splits into two parts, one interacting with the remaining scattering process via the exchange of $\tilde p_1$ and the other being a final-state parton with momentum $p_3$ propagating in the direction of $p_1$, we introduce the variable 
\beq
\label{eq:xdef}
x=\frac{1}{z}.
\eeq
In this case, $(1-x)$ can be interpreted as the fraction of momentum of $p_1$ carried forward by $p_3$ in the collinear limit.

\subsection{Regions}
\label{sec:regions_in_collinear_limit}

We now consider the asymptotic expansion around the collinear limit using the method of regions.
To this end, a key step is to determine the list of regions involved in the expansion. In this subsection, we shall explore those regions corresponding to the expansion around the time-like collinear limit, namely, $p_1$ and $p_3$ simultaneously belong to the final (or initial) state. Note that results of the space-like collinear limit can be obtained from those of the time-like collinear limit through analytic continuation. This will be discussed in section \ref{sec:Analyticcont}.

Let us start our region analysis with a more general scenario, where an arbitrary number of final-state massless partons simultaneously approach the same lightcone. For convenience, we categorize the set of external momenta into the following types:
\begin{itemize}
    \item [(1)] $L$ off-shell external momenta $q_1,\dots,q_L$;
    \item [(2)] $K$ massless external momenta $p_1,\dots,p_K$, with a set $C\subseteq \{1,\dots,K\}$, such that
    \begin{itemize}
        \item all the momenta $p_i$ ($i\in C$) are close to the same lightcone;
        \item the remaining momenta $p_i$ ($i\notin C$) are close to distinct lightcones, respectively.
    \end{itemize}
\end{itemize}
In other words, the set $C$ labels those external momenta that are collinear to each other. The corresponding asymptotic expansion, which we refer to as the \emph{collinear expansion}, is defined around the limit where all the $p_i$ are strictly on shell, and the angle between any two three-momenta $\boldsymbol{p}_{i_1}$ and $\boldsymbol{p}_{i_2}$ ($i_1,i_2\in C$) vanish at the \emph{same} speed. Namely,
\begin{subequations}
\label{eq:collinear_expansion_definition}
    \begin{align}
        & p_i^2=0 \ \ (i=1,\dots,K),\qquad q_j^2\sim Q^2\ \ (j=1,\dots,L),\\
        & p_i\cdot q_j \sim Q^2\qquad\ \forall\ i, j,\\
        & p_{i_1}\cdot p_{i_2} \sim \lambda Q^2 \quad\forall\ i_1, i_2\in C,\\
        & p_{i_1}\cdot p_{i_2} \sim Q^2 \phantom{\lambda}\quad\textup{otherwise},
    \end{align}
\end{subequations}
where $\lambda\to 0$ is the scaling parameter, and $Q$ denotes the hard scale of the scattering process.

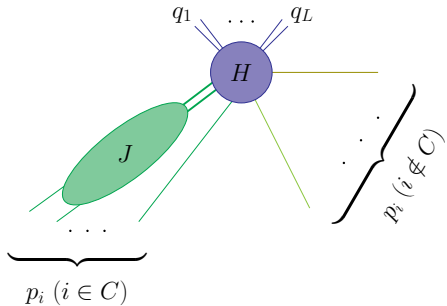
\begin{figure}[t]
\centering
\resizebox{0.4\textwidth}{!}{
\begin{tikzpicture}[baseline=11ex, line width = 0.6, font=\huge, mydot/.style={circle, fill, inner sep=.7pt}]
% \draw [help lines] (0,0) grid (10,10);

\path (6,8.2) edge [ultra thick, double, double distance=5pt, color=Green] (2.6,5.6) {};

\node (q1) at (4.5,9.5) {};
\node (q1p) at (4.7,9.7) {};
\node (qn) at (7.5,9.5) {};
\node (qnp) at (7.3,9.7) {};
\draw (q1) edge [color=Blue] (6,8) node [] {};
\draw (q1p) edge [color=Blue] (6,8) node [left] {$q_1$};
\draw (qn) edge [color=Blue] (6,8) node [] {};
\draw (qnp) edge [color=Blue] (6,8) node [right] {$q_L$};

\draw (-0.2,3.9) edge [color=Green] (2.4,5.8) node [] {};
\draw (0.6,3.6) edge [color=Green] (2.9,5.3) node [] {};
\draw (6,7.5) edge [color=Green] (3,3.6) node [] {};
\draw (8,4) edge [color=LimeGreen] (6,8) node [below] {};
\draw (10,8) edge [color=olive] (6,8) node [right] {};

\node[draw=Blue,circle,minimum size=1.8cm,fill=Blue!50] () at (6,8){};
\node[draw=Green,ellipse,minimum height=4.5cm, minimum width=1.5cm,fill=Green!50,rotate=-52] () at (2.6,5.6){};
\node at (6,8) {$H$};
\node at (2.6,5.6) {$J$};
\node at (1.2,1.5) {$p_i$ ($i\in C$)};
\node at (11,5) [rotate = 60] {$p_i$ ($i\notin C$)};

\path (8,4)-- node[mydot, pos=.333] {} node[mydot] {} node[mydot, pos=.666] {}(11,8);
\path (0,3.33)-- node[mydot, pos=.333] {} node[mydot] {} node[mydot, pos=.666] {}(3,3.33);
\path (5,9.5)-- node[mydot, pos=.333] {} node[mydot] {} node[mydot, pos=.666] {}(7,9.5);

\node () at (1.2,2.5) [rotate=0] {\huge $\underbrace{\phantom{\qquad\qquad\qquad}}$};
\node () at (10,5.5) [rotate=60] {\huge $\underbrace{\phantom{\qquad\qquad\qquad\quad}}$};

\end{tikzpicture}
}
\vspace{-2em}
\caption{The general structure of a region $R$ in the collinear expansion where all the external collinear momenta are collinear to $p_1$. Each internal propagator belongs to either the hard subgraph $H$ or the unique jet subgraph $J$, whose momenta are characterized by eq.~(\ref{eq:hard_and_collinear_scalings}).}
\label{collinear_expansion_regions_general_configuration}
\end{figure}

In general, regions that are relevant for a certain asymptotic expansion can be obtained via the \emph{Newton polytope approach} in parametric representation~\cite{PakSmn11,SmnvSmnSmv19}. For the collinear expansion described above, all the regions follow the configuration in figure~\ref{collinear_expansion_regions_general_configuration}.
Each internal edge/vertex belongs to either the hard subgraph $H$ or the jet $J_1$, both being connected subgraphs of the entire Feynman graph $G$. Two or more external momenta from the set $\{p_i\ |\ i\in C\}$ attach to $J$, meanwhile all the remaining external momenta attach to $H$ directly. In addition, all the loop momenta in $H$ and $J_1$ are in the hard mode and the collinear mode, respectively, with scaling
\begin{eqnarray}
\label{eq:hard_and_collinear_scalings}
\textup{hard mode: } k_H^\mu\sim Q(1,1,1),\qquad \textup{collinear mode: } k_C^\mu\sim Q(1,\lambda,\lambda^{1/2}),
\end{eqnarray}
with $\lambda\to 0$ the same as eq.~(\ref{eq:collinear_expansion_definition}). In the expression above, we have used the lightcone coordinate along the collinear direction: each vector $v^\mu$ is written in the form of
\[
v^\mu = (v\cdot\overline{\beta}, v\cdot\beta,v\cdot\beta_{\perp}),
\]
with $\beta^\mu$ any lightlike vector collinear to $p_i^\mu$ ($i\in C$), $\overline{\beta}^\mu$ lightlike and in the opposite direction of $\beta^\mu$, $\beta_{\perp}^\mu$ transverse to both $\beta^\mu$ and $\overline{\beta}^\mu$, and additionally, $\beta\cdot\overline{\beta} = 1$. Based on this knowledge of regions, we implemented a graph-finding algorithm to obtain the entire list of regions, equivalent to the traditional Newton polytope approach meanwhile circumventing the need for constructing Newton polytopes.

Before we investigate more properties of these regions, it is worth noting a key restriction of the Newton polytope approach: it guarantees to capture all the regions when there is a choice of kinematic regime in which all the independent Mandelstam invariants have the same sign\footnote{Equivalently, this corresponds to the scenario where the second Symanzik polynomial $\mathcal{F}$ can be chosen positive or negative definite.}, or when the kinematic region can be reached via analytic continuation from such a same-sign regime. Regions in the output of the Newton polytope approach correspond to the endpoint singularities in the Feynman parameter space, and can be characterized by the ``lower facets'' (certain codimension-$1$ faces) of the Newton polytope. For certain asymptotic expansions, there are additional regions corresponding to pinch singularities in parameter space. Such regions possibly arise when cancellation within the second Symanzik polynomial $\mathcal{F}$ coincides with some solution of the Landau equations, thus they are ``hidden'' within the interior, rather than the facets, of the original polytope. To identify such \emph{hidden regions} through the Newton polytope approach, one needs to properly change the integration variables or dissect the original polytope into distinct sectors in advance~\cite{JtzSmnSmn12,Ananthanarayan:2018tog,GrdHzgJnsMa24}.

In the context of the N$^3$LO splitting amplitudes as we center on in this paper, our approach is to consider only the regions in the time-like splitting process, i.e., the collinear external momenta are all in the initial (or final) state. Such kinematics fulfill the condition that all the independent Mandelstam invariants are of the same sign, thus figure~\ref{collinear_expansion_regions_general_configuration}, as a result of the Newton polytope approach, suffices to characterize all the regions responsible for the collinear expansion defined in eq.~(\ref{eq:collinear_expansion_definition}). Based on these regions we will first compute the time-like splitting amplitudes, which can be further extended to the space-like splitting amplitudes via analytic continuation (see section~\ref{sec:Analyticcont}). Note that as an alternative approach, one can also start from analyzing the regions in the space-like collinear limit directly, in which case the Mandelstam invariants are not of the same sign, and cancellations within $\mathcal{F}$ are possible. Nevertheless, we conjecture that no hidden regions are involved in the result (namely, figure~\ref{collinear_expansion_regions_general_configuration} still captures all the regions) because no cancellation within $\mathcal{F}$ for general kinematics (i.e. for arbitrary values of independent Mandelstam invariants) would be compatible with the Landau equations. Examining this statement and comparing the results from these two approaches could be an interesting topic, which we leave for future investigation.
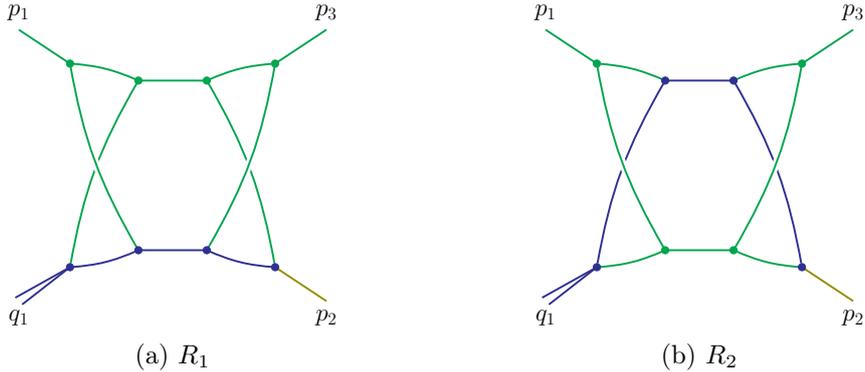
\begin{figure}[t]
\centering
\begin{subfigure}[b]{0.45\textwidth}
    \centering
    \resizebox{0.67\textwidth}{!}{
\begin{tikzpicture}[line width = 0.6, font=\large, mydot/.style={circle, fill, inner sep=.7pt}]
% \draw [help lines] (0,0) grid (10,10);

\draw (0.4,1.1) edge [ultra thick, Blue] (2,2) node [] {};
\draw (0.6,0.9) edge [ultra thick, Blue] (2,2) node [] {};
\draw (9.5,1) edge [ultra thick, olive] (8,2) node [] {};
\draw (0.5,9) edge [ultra thick, Green] (2,8) node [] {};
\draw (9.5,9) edge [ultra thick, Green] (8,8) node [] {};
\draw (4,2.5) edge [ultra thick, Blue] (6,2.5) node [] {};
\draw (4,7.5) edge [ultra thick, Green] (6,7.5) node [] {};
\draw (4,7.5) edge [ultra thick, bend right = 10, Green] (2,2) node [] {};
\draw (6,7.5) edge [ultra thick, bend left = 10, Green] (8,2) node [] {};
\draw (4,7.5) edge [ultra thick, bend right = 10, Green] (2,8) node [] {};
\draw (6,7.5) edge [ultra thick, bend left = 10, Green] (8,8) node [] {};
\draw (4,2.5) edge [ultra thick, draw=white, double=white, double distance=3pt, bend left = 10] (2,2) node [] {};\draw (4,2.5) edge [ultra thick, bend left = 10, Blue] (2,2) node [] {};
\draw (6,2.5) edge [ultra thick, draw=white, double=white, double distance=3pt, bend right = 10] (8,2) node [] {};\draw (6,2.5) edge [ultra thick, bend right = 10, Blue] (8,2) node [] {};
\draw (4,2.5) edge [ultra thick, draw=white, double=white, double distance=3pt, bend left = 10] (2,8) node [] {};\draw (4,2.5) edge [ultra thick, bend left = 10, Green] (2,8) node [] {};
\draw (6,2.5) edge [ultra thick, draw=white, double=white, double distance=3pt, bend right = 10] (8,8) node [] {};\draw (6,2.5) edge [ultra thick, bend right = 10, Green] (8,8) node [] {};

\node () at (0.5,0.5) {\huge $q_1$};
\node () at (0.5,9.5) {\huge $p_1$};
\node () at (9.5,9.5) {\huge $p_3$};
\node () at (9.5,0.5) {\huge $p_2$};

\draw[fill, thick, Blue] (2,2) circle (3pt);
\draw[fill, thick, Blue] (8,2) circle (3pt);
\draw[fill, thick, Green] (2,8) circle (3pt);
\draw[fill, thick, Green] (8,8) circle (3pt);
\draw[fill, thick, Green] (4,7.5) circle (3pt);
\draw[fill, thick, Green] (6,7.5) circle (3pt);
\draw[fill, thick, Blue] (4,2.5) circle (3pt);
\draw[fill, thick, Blue] (6,2.5) circle (3pt);

\end{tikzpicture}
}
    \vspace{-3em}\caption{$R_1$}
    \label{maximally_collinear_region1}
\end{subfigure}
\begin{subfigure}[b]{0.45\textwidth}
    \centering
    \resizebox{0.67\textwidth}{!}{
\begin{tikzpicture}[line width = 0.6, font=\large, mydot/.style={circle, fill, inner sep=.7pt}]
% \draw [help lines] (0,0) grid (10,10);

\draw (0.4,1.1) edge [ultra thick, Blue] (2,2) node [] {};
\draw (0.6,0.9) edge [ultra thick, Blue] (2,2) node [] {};
\draw (9.5,1) edge [ultra thick, olive] (8,2) node [] {};
\draw (0.5,9) edge [ultra thick, Green] (2,8) node [] {};
\draw (9.5,9) edge [ultra thick, Green] (8,8) node [] {};
\draw (4,2.5) edge [ultra thick, Green] (6,2.5) node [] {};
\draw (4,7.5) edge [ultra thick, Blue] (6,7.5) node [] {};
\draw (4,7.5) edge [ultra thick, bend right = 10, Blue] (2,2) node [] {};
\draw (6,7.5) edge [ultra thick, bend left = 10, Blue] (8,2) node [] {};
\draw (4,7.5) edge [ultra thick, bend right = 10, Green] (2,8) node [] {};
\draw (6,7.5) edge [ultra thick, bend left = 10, Green] (8,8) node [] {};
\draw (4,2.5) edge [ultra thick, draw=white, double=white, double distance=3pt, bend left = 10] (2,2) node [] {};\draw (4,2.5) edge [ultra thick, bend left = 10, Green] (2,2) node [] {};
\draw (6,2.5) edge [ultra thick, draw=white, double=white, double distance=3pt, bend right = 10] (8,2) node [] {};\draw (6,2.5) edge [ultra thick, bend right = 10, Green] (8,2) node [] {};
\draw (4,2.5) edge [ultra thick, draw=white, double=white, double distance=3pt, bend left = 10] (2,8) node [] {};\draw (4,2.5) edge [ultra thick, bend left = 10, Green] (2,8) node [] {};
\draw (6,2.5) edge [ultra thick, draw=white, double=white, double distance=3pt, bend right = 10] (8,8) node [] {};\draw (6,2.5) edge [ultra thick, bend right = 10, Green] (8,8) node [] {};

\node () at (0.5,0.5) {\huge $q_1$};
\node () at (0.5,9.5) {\huge $p_1$};
\node () at (9.5,9.5) {\huge $p_3$};
\node () at (9.5,0.5) {\huge $p_2$};

\draw[fill, thick, Blue] (2,2) circle (3pt);
\draw[fill, thick, Blue] (8,2) circle (3pt);
\draw[fill, thick, Green] (2,8) circle (3pt);
\draw[fill, thick, Green] (8,8) circle (3pt);
\draw[fill, thick, Blue] (4,7.5) circle (3pt);
\draw[fill, thick, Blue] (6,7.5) circle (3pt);
\draw[fill, thick, Green] (4,2.5) circle (3pt);
\draw[fill, thick, Green] (6,2.5) circle (3pt);

\end{tikzpicture}
}
    \vspace{-3em}\caption{$R_2$}
    \label{maximally_collinear_region2}
\end{subfigure}
\caption{An example where two maximally collinear regions, $R_1$ and $R_2$, are relevant in the collinear expansion for the same three-loop graph. Here, in line with our parameterization of the collinear limit (see section~\ref{sec:Parameterizing}), the momentum $q_1$ is off shell, $p_1$ and $p_3$ are collinear to each other, while $p_2$ is in another direction. For each region, we have colored the jet propagators in {\color{Green}\bf green} and hard propagators in {\color{Blue}\bf blue}, respectively.}
\label{figure-maximally_collinear_regions_example}
\end{figure}

Now let us return to the regions of figure~\ref{collinear_expansion_regions_general_configuration}. Among all these regions, we identify that the contributions to the N$^3$LO splitting amplitudes are provided by those \emph{maximally collinear regions}. A maximally collinear region is defined as any configuration of figure~\ref{collinear_expansion_regions_general_configuration} where all the loop momenta are in the collinear mode in eq.~(\ref{eq:hard_and_collinear_scalings}). The hard subgraph $H$ is then a connected tree graph. It is worth noting that an individual Feynman graph may feature multiple maximally collinear regions (see figure~\ref{figure-maximally_collinear_regions_example} as an example, where $p_1\parallel p_3$, and two maximally collinear regions are manifested). Once we identify the entire list of maximally collinear regions, we can obtain the leading power contribution to the collinear expansion by performing a simple Laurent series expansion in $\lambda$ at the integrand level, taking the leading terms, and finally setting $\lambda=1$.

\begin{figure}[t]
\centering
\begin{subfigure}[b]{0.45\textwidth}
    \centering
    \resizebox{\textwidth}{!}{
\begin{tikzpicture}[baseline=11ex, line width = 0.6, font=\huge, mydot/.style={circle, fill, inner sep=.7pt}]
% \draw [help lines] (0,0) grid (10,10);

\path (6,8) edge [double,double distance=2pt,color=Green] (2,5) {};
\path (6,8) edge [double,double distance=2pt,color=LimeGreen] (6,3.5) {};
\path (6,8) edge [double,double distance=2pt,color=olive] (10,5) {};

\draw (3,2) edge [dashed,double,color=Rhodamine] (6,8) node [right] {};
\draw (3,2) edge [dashed,double,color=Rhodamine,bend left = 15] (2,5) {};
\draw (3,2) edge [dashed,double,color=Rhodamine,bend right = 15] (6,3.5) {};
\draw (3,2) edge [dashed,double,color=Rhodamine,bend right = 50] (10,5) {};

\node (q1) at (4.5,9.5) {};
\node (q1p) at (4.7,9.7) {};
\node (qn) at (7.5,9.5) {};
\node (qnp) at (7.3,9.7) {};
\draw (q1) edge [color=Blue] (6,8) node [] {};
\draw (q1p) edge [color=Blue] (6,8) node [left] {$q_1$};
\draw (qn) edge [color=Blue] (6,8) node [] {};
\draw (qnp) edge [color=Blue] (6,8) node [right] {$q_L$};

\node (pn) at (12,3.5) {};
\draw (0,3.8) edge [color=Green] (2,5.3) node [] {};
\draw (0.3,3.3) edge [color=Green] (2.3,4.8) node [] {};
\draw (5.7,1) edge [color=LimeGreen] (5.7,3.5) node [] {};
\draw (6.3,1) edge [color=LimeGreen] (6.3,3.5) node [] {};
\draw (pn) edge [color=olive] (10,5) node [below] {$p_K$};

\path (6,3.5)-- node[mydot, pos=.333] {} node[mydot] {} node[mydot, pos=.666] {}(10,5);
\path (q1)-- node[mydot, pos=.333] {} node[mydot] {} node[mydot, pos=.666] {}(qn);
\path (-0.5,3.6)-- node[mydot, pos=.333] {} node[mydot] {} node[mydot, pos=.666] {}(1,2);
\path (6,0.5)-- node[mydot, pos=.3] {} node[mydot] {} node[mydot, pos=.7] {}(7.2,0.5);

\node[draw=Blue,circle,minimum size=1.6cm,fill=Blue!50] () at (6,8){};
\node[dashed, draw=Rhodamine,circle,minimum size=2.4cm,fill=Rhodamine!50] () at (3,2){};
\node[draw=Green,ellipse,minimum height=3cm, minimum width=1.1cm,fill=Green!50,rotate=-52] () at (2,5){};
\node[draw=LimeGreen,ellipse,minimum height=3cm, minimum width=1.1cm,fill=LimeGreen!50] () at (6,3.5){};
\node[draw=olive,ellipse,minimum height=3cm, minimum width=1.1cm,fill=olive!50,rotate=52] () at (10,5){};

\node at (6,8) {$H$};
\node at (3,2) {$S$};
\node at (2,5) {$J_1$};
\node at (6,3.5) {$J_2$};
\node at (10,5) {$J_K$};
\node at (-0.4,3.6) {$p_1$};
\node at (5.7,0.5) {$p_2$};

\end{tikzpicture}
}
\vspace{-2em}
    \captionsetup{width=0.9\linewidth}\caption{Regions with multiple jets and a nonempty soft subgraph between them.}
    \label{collinear_expansion_regions_multiple_jets}
\end{subfigure}
\qquad
\begin{subfigure}[b]{0.45\textwidth}
    \centering
    \resizebox{0.8\textwidth}{!}{
\begin{tikzpicture}[baseline=11ex, line width = 0.6, font=\huge, mydot/.style={circle, fill, inner sep=.7pt}]
% \draw [help lines] (0,0) grid (10,10);

\path (1.5,7.5) edge [color=Green] (3,7.5) {};
\path (8.5,7.5) edge [color=Green] (7,7.5) {};
\path (1.5,2.5) edge [color=LimeGreen] (3,2.5) {};
\path (8.5,2.5) edge [color=LimeGreen] (7,2.5) {};
\path (3.6,5) edge [color=Blue] (4.8,5.1) {};
\path (3.6,5) edge [color=Blue] (4.8,4.9) {};
\path (3,7.5) edge [very thick, double, double distance=2pt,color=Green] (3.6,5) {};
\path (3,2.5) edge [very thick, double, double distance=2pt,color=LimeGreen] (3.6,5) {};

\draw (7,5) edge [dashed,double,color=Orange,bend left = 30] (5,2.5) {};
\draw (7,5) edge [dashed,double,color=Orange,bend right = 30] (5,7.5) {};

\node[draw=Blue,circle,minimum size=1.2cm,fill=Blue!50] () at (3.6,5){};
\node[dashed, draw=Orange,circle,minimum size=2.4cm,fill=Orange!50] () at (7,5){};
\node[draw=Green,ellipse,minimum height=5cm, minimum width=1.1cm,fill=Green!50,rotate=90] () at (5,7.5){};
\node[draw=LimeGreen,ellipse,minimum height=5cm, minimum width=1.1cm,fill=LimeGreen!50,rotate=90] () at (5,2.5){};

\node at (3.6,5) {$H$};
\node at (7,5) {Glauber};
\node at (5,7.5) {$J_1$};
\node at (5,2.5) {$J_2$};
\node at (1,7.5) {$p_1$};
\node at (9,7.5) {$p_2$};
\node at (1,2.5) {$p_3$};
\node at (9,2.5) {$p_4$};
\node at (5.2,5) {$q_1$};

\end{tikzpicture}
}
\vspace{-2em}
    \caption{Regions with momenta in the Glauber mode exchanging between two jets.}
    \label{collinear_expansion_regions_Glauber}
\end{subfigure}
\caption{Regions involved in some more general collinear expansions, where (a) two or more sets of external momenta are collinear to distinct directions ($p_1$ and $p_2$ in this example), respectively, and (b) multiple space-like splittings in distinct directions are involved. In the case of (a) two or more nontrivial jets and an nonempty soft subgraph $S$ can emerge, while in (b) there can be a subgraph adjacent to both spectator jets, carrying loop momenta in the Glauber mode.}
\end{figure}
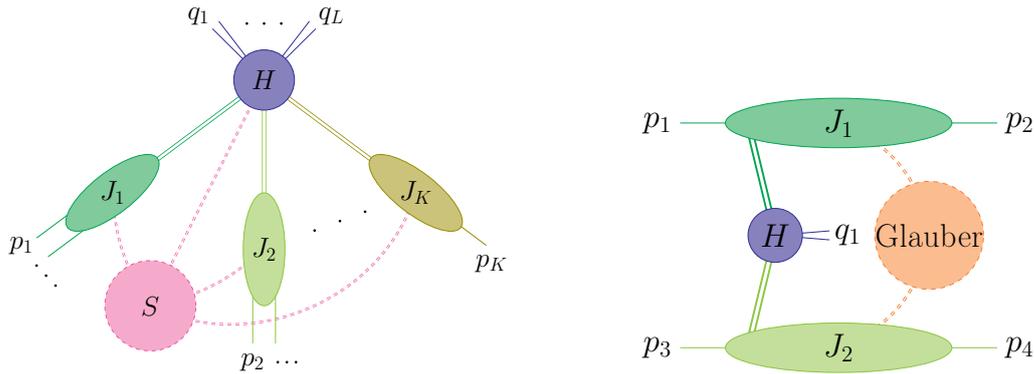

To end this section, we discuss the potential extension of our approach to incorporate more general collinear expansions.
Note that for more general collinear expansions involving multiple sets of collinear momenta in distinct directions, two or more nontrivial jets, and a nonempty soft subgraph with all its loop momenta scaling as $Q(\lambda,\lambda,\lambda)$, can appear in the regions. This scenario is illustrated in figure~\ref{collinear_expansion_regions_multiple_jets}, with further requirements of the hard, jet, and soft subgraphs needed~\cite{Ma23}. Complexity increases further with multiple space-like splittings, potentially involving the Glauber mode $Q(\lambda,\lambda,\lambda^{1/2})$. For example, in the Drell-Yan process shown in figure~\ref{collinear_expansion_regions_Glauber}, propagators with Glauber-mode momenta can exchange between the two spectators. We shall defer these interesting generalizations to future studies.
\section{Splitting amplitudes}
\label{sec:splittings}
%\subsection{QCD splitting amplitudes}
\subsection{QCD}
\label{sec:calculation}

We derive a general parameterization of the scattering amplitudes listed in eq.~\eqref{eq:amplist} in conventional dimensional regularization and subsequently project the amplitudes on a spinor-helicity basis in the 't Hooft-Veltman scheme, treating only external polarizations as four dimensional.
We use QGRAF~\cite{qgraf} to generate Feynman graphs and private software to perform color, spinor, and Lorentz algebra to obtain a Feynman integrand for the scattering amplitudes.
We then use the parameterization introduced in section \ref{sec:Parameterizing} in conjunction with the collinear expansion techniques outlined in section \ref{sec:regions_in_collinear_limit} to extract the leading term in the limit $\lambda \to 0$ of the amplitudes.
After expansion, we identify suitable integrand topologies and reduce the collinear loop integrals to a set of 553 collinear master integrals using IBP identities~\cite{Tkachov1981,Laporta:2001dd,Chetyrkin1981} via the tool \texttt{Blade}~\cite{Guan:2024byi} and another custom implementation of the Laporta algorithm~\cite{Laporta:2001dd}. 
We compute our master integrals using the method of differential equations~\cite{Kotikov:1991pm,Gehrmann:1999as,Kotikov:1990kg,Kotikov:1991hm,Henn:2013pwa} using algorithmic techniques~\cite{Lee:2014ioa}.
We compute boundary conditions using regularity and consistency conditions (see, for example, refs.~\cite{Henn:2020lye,Dulat:2014mda,Henn:2013nsa}).
Note that a subset of master integrals was already computed for the purposes of ref.~\cite{Herzog:2023sgb}.
Dividing our amplitudes by the corresponding Born amplitudes yields the desired splitting amplitudes as per eq.~\eqref{eq:splitformula}.
We derive bare, unrenormalized scattering amplitudes but anticipate renormalization in the $\overline{\text{MS}}$ scheme by absorbing ubiquitous factors in the definition of the parameter $a_S$ introduced in eq.~\eqref{eq:asdef}. 
This parameter is related to its bare counterpart via
\beq
a_S=a_S^0 \left(\frac{4\pi}{\mu^2}\right)^{\epsilon}e^{-\epsilon \gamma_E},
\eeq
with $\gamma_E=0.577215664902$ the Euler-Mascheroni constant. 
We work in dimensional regularization such that the spacetime dimension $d$ is related to the dimensional regulator via $d=4-2\epsilon$.

Below, we parameterize the splitting amplitudes in terms of the Born-level scattering amplitude multiplied by a perturbative, scalar factor for each splitting process and helicity configuration.
These scalar functions depend logarithmically on the ratio $s_{13}/\mu^2$ and, via harmonic polylogarithms~\cite{Remiddi:1999ew} and rational functions, on the variable $z$. 
We use in-house implementations of algorithms derived in refs.~\cite{Duhr:2012fh,Duhr:2011zq,Goncharov:2010jf,Huber:2005yg,Duhr:2019tlz,Panzer:2014gra} to manipulate these polylogarithms.
We present our results for these scalar functions in terms of electronically readable ancillary files attached to the arXiv submission of this article.
The functions also depend on the number of massless quark flavors $n_f$ as well as on the Casimir invariants of $SU(n_c)$, where $n_c=3$ for QCD:
\beq
C_A=n_c,\hspace{1cm} C_F=\frac{n_c^2-1}{2n_c},
\eeq
\beq
C_4^{AA}=\frac{n_c^2}{24} (n_c^2-1)(36+n_c^2),\quad
C_4^{AF}=\frac{n_c}{48} (n_c^2-1)(6+n_c^2),\quad
C_4^{FF}=\frac{(n_c^2-1)(18-6n_c^2+n_c^4)}{96n_c^2} .
\eeq
%We also like the following variables
%\beq
%w=\frac{s_{13}}{s_{12}}=-\lambda\frac{k_\perp^2}{z^2(1-z)},\hspace{1cm} \bar w=\frac{s_{23}}{s_{12}}=\frac{1-z}{z}.
%\eeq
%Most important is that only $\bar w$ becomes small in the collinear limit as $\lambda \to 0$.
For a generic representation the invariants $C_4^{R_1 R_2}$ are defined in terms of traces of the generators $T^a_R$ of the group representations via 
\beq
C_4^{R_1R_2} = d_{R_1}^{abcd} d_{R_2}^{abcd},\hspace{1cm}d_R^{abcd}= \frac{1}{4!} \left[\text{Tr}\big(T_R^aT_R^bT_R^cT_R^d\big) +\text{symmetric permutations}\right]\,.
\eeq

\subsection*{$\boldsymbol{g\to gg}$ splitting amplitudes}
\label{sec:ggg}
The splitting amplitudes of a gluon splitting to two gluons
 are given by
\begin{subequations}
\begin{align}
    \text{Split}_{-}^{a c_1 c_2}(\{p_1,g,+\},\{p_3,g,+\})&=\frac{\sqrt{2}g_Sf^{a c_1 c_2}}{\sqrt{z(1-z)} \langle 1 | 3 \rangle } S_{g+\to g+g+}(z) \\
    \text{Split}_{-}^{a c_1 c_2}(\{p_1,g,+\},\{p_3,g,-\})&=\frac{\sqrt{2} g_S f^{a c_1 c_2}}{\sqrt{z(1-z) }[ 3 | 1 ]}z^2 S_{g+\to g+g-}(z). \\
    \text{Split}_{-}^{a c_1 c_2}(\{p_1,g,-\},\{p_3,g,+\})&=\frac{\sqrt{2}g_S f^{a c_1 c_2}}{\sqrt{z(1-z)}[3|1]}(1-z)^2 S_{g+\to g-g+}(z). \\
    \text{Split}_{-}^{a c_1 c_2}(\{p_1,g,-\},\{p_3,g,-\})&=-\frac{\sqrt{2} g_Sf^{a c_1 c_2} \langle 1 | 3 \rangle}{\sqrt{z(1-z) } [ 3 | 1 ]^2 }  z(1-z) S_{g+\to g-g-}(z).
\end{align}
\end{subequations}
Above, $f^{abc}$ is a structure constant of $SU(n_c)$.\footnote{Note, that we used the identity $d_{4R^\prime}^{abcd}T_{R,\,ij}^bT_{R,\,jk}^cT_{R,\,kl}^d=\frac{C_4^{RR^\prime}}{T_R d_A} T_{R,\, il}^a$, which is valid for $SU(n_c)$, to express our splitting amplitudes in the form outlined here.}
The functions $S_X$ can be expanded perturbatively,
\beq
\label{eq:splitfacexpansion}
S_X=\sum_o a_S^o S_X^{(o)},
\eeq
and at the leading order we find
\beq
S^{(0)}_{g+\to g+g+} =S^{(0)}_{g+\to g+g-} =S^{(0)}_{g+\to g-g+} =1, \hspace{1cm} S^{(0)}_{g+\to g-g-}=0. 
\eeq

The remaining amplitudes ($\lambda_P =-$) can be found simply by performing the following replacement
\beq
\label{eq:helswap}
\text{Split}_{-(-\lambda_P)}^{a c_1 c_2}(\{p_1,g,\lambda_1\},\{p_3,g,\lambda_3\})=-\text{Split}_{-\lambda_P}^{a c_1 c_2}(\{p_1,g,-\lambda_1\},\{p_3,g,-\lambda_3\})\Bigg|_{\langle a b\rangle\leftrightarrow [ab]}.
\eeq

We observe that the functions $S_X$ satisfy certain symmetries under the exchange of $z\leftrightarrow 1-z$:
\begin{subequations}
    \begin{align}
        S_{g+\to g+g+}(1-z)=&S_{g+\to g+g+}(z),\\
        S_{g+\to g+g-}(1-z)=&S_{g+\to g-g+}(z),\\
        S_{g+\to g-g+}(1-z)=&S_{g+\to g+g-}(z),\\
        S_{g+\to g-g-}(1-z)=&S_{g+\to g-g-}(z).
    \end{align}
\end{subequations}

\subsection*{$\boldsymbol{g \to q\bar q}$ splitting amplitudes}
\label{sec:gqqbar}
The splitting amplitudes for a gluon to split into a quark-anti-quark pair are given by
\bea
\text{Split}_{-}^{a c_1 c_3}(\{p_1,q,-\},\{p_3,\bar q,+\})&=&- \frac{z\sqrt{2} g_S T^{a}_{c_3 c_1}}{  [ 3 | 1 ]}S_{g^+\to q^- \bar q^+}(z), \nonumber\\
\text{Split}_{-}^{a c_1 c_3}(\{p_1,q,+\},\{p_3,\bar q,-\})&=&- \frac{ (1-z)\sqrt{2} g_S T^{a}_{c_3 c_1}}{  [ 3 | 1 ] }S_{g^+\to q^+ \bar q^-}(z). 
\eea
The scalar splitting factors $S_X$ can be expanded perturbatively as in eq.~\eqref{eq:splitfacexpansion} and at Born-level we find.
\beq
S^{(0)}_{g^+\to q^+ \bar q^-}=S^{(0)}_{g^+\to q^- \bar q^+}=1
\eeq
As above, the $\lambda_P =-$ helicity amplitudes can be obtained using eq.~\eqref{eq:helswap}. Furthermore, we observe that the functions transform into each other under the exchange of $z\to 1-z$.
\beq
S_{g^+\to q^- \bar q^+}(1-z)=S_{g^+\to q^+ \bar q^-}(z).
\eeq

\subsection*{$\boldsymbol{q\to q g}$ splitting amplitudes}
\label{sec:qqg}
The splitting amplitudes for a quark to split into a quark and a gluon are given by
\begin{subequations}
    \begin{align}
        \text{Split}_{-}^{a c_1 c_3}(\{p_1,q,+\},\{p_3,g,+\})&=\frac{ \sqrt{2} g_S T^{c_3}_{c_1 a}}{\sqrt{1-z} \langle 1 | 3\rangle }S_{q^+\to q^+ g^+}(z), \\
        \text{Split}_{-}^{a c_1 c_3}(\{p_1,q,+\},\{p_3,g,-\})&=\frac{ z\sqrt{2} g_S T^{c_3}_{c_1 a}}{\sqrt{1-z}[ 3 | 1 ] }S_{q^+\to q^+ g^-}(z).
    \end{align}
\end{subequations}
The scalar splitting factors $S_X$ can be expanded perturbatively as in eq.~\eqref{eq:splitfacexpansion} and at Born-level we find
\beq
S^{(0)}_{q^+\to q^+ g^+}=S^{(0)}_{q^+\to q^+ g^-}=1\,.
\eeq
As above, the $\lambda_P =-$ helicity amplitudes can be obtained using eq.~\eqref{eq:helswap}.

%\subsection{$\mathcal{N}=4$ sYM splitting amplitudes}
\subsection{$\mathcal{N}=4$ sYM}
\label{sec:n4sym}

Maximally supersymmetric Yang-Mills theory ($\mathcal{N}=4 $ sYM) has proven to be an excellent testing ground for many aspects of four-dimensional non-Abelian gauge theory. 
This theory serves as a laboratory to explore properties of four-dimensional gauge theory and to develop new insights that may be applied to QCD.
One remarkable observation is that there is a similarity between QCD and $\mathcal{N}=4$ sYM: the leading transcendental part of the perturbative expansion of certain quantities agrees between the two theories~\cite{Kotikov:2004er,Kotikov:2002ab}.
It has been shown that this correspondence holds true for certain form factors of operators of the stress tensor multiplet~\cite{Jin:2018fak,Brandhuber:2017bkg,Brandhuber:2014ica,Brandhuber:2012vm}.
The form factor of three on-shell states $\Phi$ and the trace of two scalar fields $\phi$,
\beq
\mathcal{F}_2=\int d^d x \langle \Phi_1 \Phi_2\Phi_3 | \phi^I(x) \phi^I(x) |0\rangle,
\eeq
corresponds to the amplitude of a Higgs boson decaying to three gluons in QCD. 
This form factor has been studied in many contexts by the community~\cite{Dixon:2022xqh,Dixon:2021tdw,Dixon:2020bbt,Dixon:2023kop,Yang:2019vag,Guan:2023gsz,Lin:2021qol,Lin:2021kht} and was recently obtained to fantastic eight-loop accuracy in the planar limit of the theory~\cite{Dixon:2022rse}.

Similar to the QCD case discussed above, the collinear limit of these form factors can be used to extract the splitting amplitudes in $\mathcal{N}=4$ sYM theory. 
To achieve this, our starting point is the integrand for the form factor determined in ref.~\cite{Lin:2021kht} at two- and three-loop order.
We then apply our integrand expansion technology and compute the first term in the maximally collinear limit of this form factor. 
We obtain a pure function (i.e. of maximal transcendental weight) for both the two- and three-loop result.
We then compare our result with the maximally collinear limit of the $\mathcal{A}_{hggg}$ amplitude.
First, we observe that the leading transcendental part of $S_{g+\to g-g-}$ vanishes. 
Second, we observe that the leading transcendental part of all other helicity amplitudes is identical.
Third, we find perfect agreement with the new result of the strict collinear limit of the $\mathcal{N}=4$ sYM theory amplitude.
Finally, we checked the leading transcendental part of all other splitting amplitudes involving quarks in the final or initial state.
When changing the color representation of quarks from fundamental to adjoint we recover again the $\mathcal{N}=4$ sYM theory result.
This validates the principle of maximal transcendentality for the splitting amplitudes computed here.
We would like to emphasize that the correspondence holds both for leading and sub-leading color contributions.

%\subsection{Analytic continuation and space-like splitting}
\subsection{Analytic continuation}
\label{sec:Analyticcont}
The collinear limit of a scattering amplitude into a true product of a splitting amplitude and a lower-multiplicity amplitude as in eq.~\eqref{eq:splitformula} is accurate if the partons resulting from the splitting process are all in the final (or initial) state, referred to as time-like splitting process.
However, in the kinematic configuration where one parton of the splitting pair is in the initial state while the other is in the final state - referred to as a space-like splitting process - a modification of eq.~\eqref{eq:splitformula} including color correlations with other partons not participating in the splitting process is necessary. This fact, referred to as the violation of strict collinear factorization, was introduced and discussed in refs.~\cite{Cieri:2024ytf,Catani:2011st,Catani:2012iw,Forshaw:2012bi,Kyrieleis:2006fh,Forshaw:2006fk,Schwartz:2017nmr}.
A collinear factorization formula correctly taking into account color correlations that arise for space-like splitting processes was introduced in ref.~\cite{Catani:2011st} and recently worked out explicitly at the two-loop order~\cite{Henn:2024qjq}.
Here, from the results for the time-like splitting process (for scattering amplitudes with arbitrarily many partons), we perform the analytic continuation explicitly for scattering amplitudes involving three external partons, and leave a discussion for more general amplitudes to future work.

We performed our computation of splitting amplitudes in a pseudo-Euclidean region in which all Lorentz invariant scalar products of parton momenta are negative.
At the $L^{\text{th}}$ loop order the splitting amplitudes, derived from our boson-and-three-parton amplitudes, take the form
\beq
\label{eq:splittingphases}
S_X^{(L)}(z,s_{13},\mu^2)=\left(\frac{-s_{13}-i 0}{\mu^2}\right)^{-L \epsilon}\sum_{i=0}^L \left(\frac{-s_{23}-i0 }{-s_{12}-i0 }\right)^{-i\epsilon}S_X^{(L,i)}(z).
\eeq
The above equation makes all branch points manifest when crossing external particles from initial to final state and the functions $S_X^{(L,i)}$ are real for $z\in [0,1] $.
In physical kinematics, we distinguish the two cases where the two partons resulting from the splitting process are in the final state (time-like) and where one is in the initial state (space-like).
We assume $p_2$ is in the initial state.
\bea
\text{Time-like splitting:}&&\text{$p_1$ and $p_3$ are both in the final state},\nonumber\\
&&\hspace{1cm}  s_{12}<0,\hspace{0.5cm} s_{13}>0,\hspace{1cm}s_{23}<0.\\
\text{Space-like splitting:}&&\text{$p_1$ is in the initial state while $p_3$ in the final state},\nonumber\\
&&\hspace{1cm}  s_{12}>0,\hspace{0.5cm} s_{13}<0,\hspace{1cm}s_{23}<0.\nonumber
\eea
The corresponding splitting amplitudes can be determined by adjusting the necessary phases in eq.~\eqref{eq:splittingphases} and expanding in the dimensional regulator.
Analytic continuation into the space-like region of the splitting factors $S_X$ depends on whether $p_2$ is in the final or initial state. 
Fortunately, when computing QCD corrections to physical processes only the modulus of the splitting factor $|S_X|$ appears and the dependence of the phase on the choice of reference vector $p_2$ drops out.
This would not neccessarily be the case for space-like splitting with more partons in the final state, where one should choose different reference vectors in different color ordered amplitudes.
In this article, we explicitly derive splitting factors in time-like and space-like kinematics and select an in-coming reference momentum. 
The time-like splitting factors are independent of this choice.
Space-like splitting factors with an out-going (instead of in-coming) reference vector $p_2$ can easily be obtained by complex conjugation of our explicit results.
Similarly, space-like splitting factors with $p_1$ in the final and $p_3$ in the initial state are obtained by complex conjugation.
As alluded to above, dependence on other scattering parton momenta is more intricate if amplitudes with higher parton multiplicity are considered.

The variable $z$ is suitable to express the time-like splitting process, as in this kinematic regime $z\in[0,1]$ and we can chose real-valued harmonic polylogarithms with argument $z$ to express our functions.
The variable $x=1/z$ introduced in eq.~\eqref{eq:xdef} is most suitable for space-like splitting amplitudes with $x\in[0,1]$ in the space-like kinematic regime. 
We can express the functions $S_X\left(\frac{1}{x}\right)$ again in terms of real-valued harmonic polylogarithms with argument $x$. 
The only branch cut that is crossed is located at $z=1$ and is made manifest explicitly by eq.~\eqref{eq:splittingphases}, with the functions $S_X^{(L,i)}(z)$ holomorphic at this point.
As a consequence, to transition from our time-like splitting amplitudes to the space-like splitting amplitudes (valid for three parton amplitudes) one can simply apply the following replacements.
\beq
\log\left(\frac{s_{13}}{\mu^2}\right) \to \log\left(-\frac{s_{13}}{\mu^2}\right)+ i \pi,\hspace{0.5cm}\log(1-z)\to \log(1-x)- \log{x}+i \pi,\hspace{0.5cm} z \to \frac{1}{x}.
\eeq
Note, that to apply the above rule all logarithms in $(1-z)$ must be made explicit, for example, via shuffle relations of the harmonic polylogarithms.

\subsection{Checks}
\label{sec:checks}
We performed the following checks to assure the reliability of our computed N$^3$LO splitting amplitudes. First, we confirm that our results are consistent through second loop order with the previous computation presented in refs.~\cite{Badger:2004uk,Bern:2004cz}. Second, we check that the infrared and ultraviolet singularities of our splitting amplitudes are correctly predicted as derived in refs.~~\cite{Almelid:2015jia,Aybat:2006mz,Aybat:2006wq,Catani:1998bh,Dixon:2008gr,Korchemsky:1987wg,Sterman:2002qn,Becher:2019avh}.
See for example section 5 of ref.~\cite{Herzog:2023sgb} for a detailed description.
Third, we confirm that the limit where one of the gluons involved in the splitting amplitudes is soft matches the prediction based on the one-emission soft current computed in refs.~\cite{Herzog:2023sgb,Chen:2023hmk}.
(This applies necessarily only to the $g\to gg$ and $q\to qg$ splitting process as the amplitude used to compute the $g\to q \bar q$ splitting process vanishes in the limit of a soft gluon).
We took the collinear limit of the planar three-loop scattering amplitudes for the decay of a heavy vector gauge boson to a pair of quarks and a gluon computed in ref.~\cite{Gehrmann:2023jyv} and find agreement with our computation of the $q\to q g$ splitting amplitude.
Finally, the fact that all leading transcendental parts of our splitting amplitudes are consistent with the maximum transcendentality principle serves as another strong cross-check, see section~\ref{sec:n4sym} for more details.
\section{Conclusions}
\label{sec:conclusions}
We present the complete computation of massless QCD splitting amplitudes through third loop order in QCD.
Splitting amplitudes are the universal building block multiplicatively relating $n$-parton scattering amplitudes to ($n+1$)-parton scattering amplitudes in the limit of two partons of the latter becoming collinear to each other. 
As such our results represent ubiquitous building blocks in multi-loop QCD computations.

The results we obtained are expressed in terms of logarithms of the virtuality of the splitting partons and depend via rational functions and harmonic polylogarithms on the variable $z$.
This variable $z$ represents the momentum fraction of the parent parton which is carried away by one of the daughter partons.
To facilitate the use of the obtained splitting amplitudes we attach them in electronically readable form to the arXiv submission of this article.

Our results represent key ingredients for a large set of applications in perturbative QFT.
First, they are ingredients for universal subtraction or slicing schemes for differential cross section computations at N$^4$LO in perturbative QCD.
They may present crucial ingredients for the resummation of large collinear logarithms. 
Furthermore, they serve as one essential part in the determination of the DGLAP evolution of parton distribution functions at N$^3$LO.
In addition, they represent crucial data for the validation or bootstrap of scattering amplitudes at third loop order. 

We confirm that the connection of the maximum transcendentality principle among QCD and $\mathcal{N}=4$ sYM theory persists for splitting amplitudes at third loop order. Remarkably, this is true for both planar and non-planar contributions. 

A long-standing puzzle in QCD cross-section computations is the violation of collinear factorization at high loop orders. Our computation provides crucial ingredients to further analyze this puzzle.
Specifically, our splitting amplitudes describe any time-like one-to-two splitting process, and we present analytic continuation for space-like processes involving three partons. While highly motivated, we leave the generalization to all-multiplicity space-like processes for future work. It will be interesting to investigate whether the regions in the collinear expansion employed here will be sufficient to address this case, or whether additional Glauber/Coulomb-type regions will be required.

\acknowledgments
We would like to thank Simon Badger for a useful correspondence and Lance Dixon for useful discussions. XG, BM and AS are supported by the United States Department of Energy, Contract DE-AC02-76SF00515. FH is supported by the UKRI FLF grant ``Forest Formulas for the LHC'' (Mr/S03479x/1) and the STFC Consolidated Grant ``Particle Physics at the Higgs Centre''. YM is supported by the Swiss National Science Foundation through its project funding scheme, grant number 10001706.
\appendix

\addcontentsline{toc}{section}{References}
\bibliographystyle{jhep}
\bibliography{refs}

\end{document}